\documentstyle[12pt]{article}

\newcommand{\half}{\frac 1 2 }
\newcommand{\eg}{{\em e.g.} }
\newcommand{\be}{\begin{eqnarray}}
\newcommand{\ee}{\end{eqnarray}}
\newcommand{\noi}{\noindent}
\newcommand{\ie}{{\it i.e. }}

\begin{document}

\centerline{\Large\bf Thermodynamics of $N=1$ supersymmetric QCD}
\vspace* {-35 mm}
\begin{flushright}  USITP-95-09 \\
October-1995\\
hep-th/9510045
\end{flushright}
\vskip 55mm
\centerline{\bf J. Grundberg$^{\dagger}$, T.H. Hansson$^{\star}$
and U. Lindstr{\"o}m$^{\star}$ }
\vskip 15mm

\centerline{\bf ABSTRACT}
\vskip 3mm
We use the dual description  proposed by Seiberg
to calculate the pressure in the low temperature confined phase
of $N=1$ supersymmetric QCD using perturbation theory to $O(g^3_m)$, where
$g_m$ is the gauge coupling in the dual theory. Combining this result with
the usual high temperature expansion based on asymptotic freedom, we
study how the physics in  the intermediate temperature
regime depends on the relative size of the scale parameters in the two
descriptions. In particular we explore the possibility of having a
temperature range where both perturbation expansions are valid.

\vfil
\noindent
$^{\dagger}$Department of Mathematics and Physics,
 M{\"a}lardalens h{\"o}gskola,\\
Box 833, S-721 23 V{\"a}ster{\aa}s, Sweden\\
\noi
$^{\star}$Fysikum, University of Stockholm, Box 6730, S-11385 Stockholm,
Sweden\\
\noi
$^{\star}$ Supported by the Swedish Natural Science
Research Council (NFR).
\eject

\newcommand \sss {\mbox{ $<\overline{s}s>$} }
\def\fk{\mbox{ $f_K$} }

\noi{\large\bf 1. Introduction }

\noi
A standard method to investigate the phase diagram for a
theory at finite temperature is to use high-temperature and low-temperature
expansions of the free energy: At each temperature one calculates the free
energy by using the variables appropriate at that temperature.
In QCD, asymptotic freedom tells us that
the high $T$ expansion can be done in perturbation theory and that the relevant
degrees of freedom are quarks and gluons. At low temperature, however,
the theory becomes strongly coupled, confinement sets in and other
variables have to be used. The low $T$ degrees of freedom are the
hadrons, and  one calculates the low $T$ expansion using a phenomenological
description of a gas of the lightest hadrons, \ie the pions.\footnote{A
good introduction to finite temperature field theory containing a list of
references
to the original work is the book by Kapusta \cite{KAPU}.}

Recently, there has been a break-through in the understanding of  the
infrared structure of certain 4D supersymmetric Yang-Mills theories.
The strongest results are for $N=2$ SUSY \cite{SW}, where the low-energy
effective
action can be calculated exactly, but there are also very interesting
results for $N=1$ \cite{SEI1, INSE}. (For an introduction and a list of
references see \cite {SEI2}).
In particular,
Seiberg has argued that there often exists two dual descriptions
of the same theory, one "electric" and one "magnetic".
As in usual electric-magnetic duality in the presence of monopoles,
strong coupling in one description
implies weak coupling in the other, and a "fundamental" field in one
description can appear as a  soliton
 in the other.

In the models  considered by Seiberg, the
duality between weak and strong coupling is manifested in a rather surprising
way: Both descriptions are SUSY gauge theories, with the same global
$SU(N_f)$ flavour symmetry, but with different gauge symmetry.
In the case of an $SU(N_c)$ color symmetry in the electric theory,
the magnetic  gauge group is $SU(N_f-N_c)$. Since the beta function is
$\sim(3N_c - N_f)$, the theories are asymptotically free (UVF) or infrared
free (IRF) depending on the value of $N_f$.
For $ N_f < \frac 3 2 N_c$, the electric theory is UVF and the magnetic IRF.
For $N_f > 3N_c$, the
reverse holds, and for $ \frac 3 2  N_c < N_f < 3N_c$, they are both UVF.
Seiberg's
duality conjecture, which is supported by several very strong consistency
checks, asserts that the two theories are identical in the infrared.
Furthermore, stability of the   ground state imposes the condition
$N_f \ge N_c$, and it turns out the the cases $N_f = N_c$ and $N_f = N_c+1$
are special \cite{SEI1}. In the following we shall restrict $N_f$ to be larger
than $N_c+1$. If the gauge group of the electric theory is $SO(N)$, the
magnetic theory has gauge group $SO(N_f - N_c +4)$ and we shall consider
the intervals $N_c <  N_f < 3(N_c -2)/2$ and $N_c >4$, or $N_f =N_c >5$,
where again the magnetic theory is IRF. It is of interest
to study both the $SU(N)$
and the $SO(N)$ case since we expect a phase transition from a low
temperature confined phase to a high temperature deconfined phase for
$SO(N)$. In  the $SU(N)$ case, no deconfining phase transition is expected
since
there is matter in the fundamental representation that will  give a
perimeter behaviour to the Wilson loop even at zero temperature
(or, equivalently, a non zero value to the Polyakov loop).

The main idea of this paper is to exploit the duality in the range
$ N_c + 1 < N_f < \frac 3 2 N_c$ (for $SU(N)$) to calculate the
leading contributions to the
pressure in the electric theory both at high and low $T$. Since this
 theory is UVF  we have the standard perturbative high $T$ expansion,
but at low energy we can now invoke duality and use perturbation theory in the
IRF magnetic theory.

In the next section we outline the calculation of the
pressure both at low and high temperature while referring some details to an
Appendix. In section 3 we discuss the range of validity of the
expansions and to what extent perturbation theory can be used to obtain
information about the finite $T$ phase structure. In particular we discuss
the importance of the scale parameters that control the electric and
magnetic coupling constants in the asymptotic regions of large and high
temperature respectively.

\vskip 3mm \noi
\noi{\large\bf 2. Calculations }

In superfield notation \cite{WESS,BOOK} the action for SUSY QCD is given by,
\be
S= \frac 1 4 \int d^4x\, d^2\,\theta Tr( W^{\alpha}W_{\alpha})+h.c.+
\int d^4xd^4\theta [(\bar{\phi}_i\phi_i)+(\tilde{\bar{\phi}}_i
\tilde{\phi}_i)]
\ee
where $W^{\alpha}=W^{\alpha A}T^A$ is the $SU(N_c)$
gauge field strength and the covariantly chiral
matter superfields $\phi^a_i$ transform according to the $\bf N_c$
representation of the gauge group ($a=1,...,N_c$ is the $SU(N_c)$ index,
$i=1,..,N_f$ is the flavour index). The covariantly chiral superfields
$\tilde\phi_{ai}$ transform according to the $\bf \overline{N_c}$
representation.
Both $\phi$ and $\tilde{\phi}$ are needed to get a non-chiral theory.
In terms of ordinary fields the theory contains $N_g = N_c^2 -1$ gluons
($A_{\mu}^A$),
and gluinos, ($\lambda^{\alpha A}$, $\bar{\lambda}^{\dot{\alpha}A}$),
$A=1,..,N_g$, and also $N_c N_f$ quarks ($\psi_{\alpha a}$,
$\bar{\psi}_{\dot{\alpha}}^a$,
$\tilde{\psi}^{\alpha b}$,
$\bar{\tilde{\psi}}^{\dot{\alpha}}_b$)  and squarks ($\varphi
_a$, $\varphi^{\star}_a$, $\tilde{\varphi}^a$, $\tilde{\varphi}^{\star
a}$),  $a=1,...,N_c$. The flavour indices on quarks and squarks are
suppressed. For $SO(N)$, the number of gluons is $N_g = N_c(N_c - 1)/2$, and
the matter fields are in a real representation.

Greens functions in SUSY theories are usually calculated using supergraph
techniques  \cite{WESS,BOOK}. At finite temperature, bosons and fermions do not
enter the calculation on equal footing, and it is better to use a component
formalism.\footnote
{Technically this difference enters through the different
boundary conditions for bosonic and fermionic fields in the imaginary, or
''temperature'' direction. This difference ''breaks the supersymmetry'' in
the sense that thermal expectaion values do not satisfy the zero
temperature supersymmetry Ward identities, as follows already
from the supersymmetry algebra \cite{GIRA}. The theory is of course still
supersymmetric since thermodynamics is determined by the ($T=0$) spectrum
which is supersymmetric.}
In the Appendix we define our conventions, and give
the component form of the Lagrangian that give the Feynman rules shown in
figs. 6 - 9.

We can now use the standard machinery of finite temperature field theory to
calculate the free energy density ${\cal F}= - P$ at temperature $T$.
The calculations in a theory with zero masses and chemical potentials are
very straightforward, as demonstrated by an example in the Appendix.

To $O(g^0)$   the pressure is due to the "blackbody"
radiation from the massless degrees of freedom.
For SUSY QCD with gauge group $SU(N_c)$,
the bosonic degrees of freedom are the $2N_g$ gluons and the
$4N_cN_f$ squarks. There is an equal number of fermionic partners to these
(gluinos
and quarks), and thus the "blackbody" part of the pressure is given by
\be
{\it P}_0^{el} &=&  (2N_g+4N_cN_f) \left(1+\frac 7 8\right)
\frac{\pi^2T^4}{90}\nonumber\\
&=&(N_c^2+2N_cN_f-1)\frac{\pi^2T^4}{24}\ \ \ \ \ .
\ee
To get the corresponding result for  $SO(N)$, one substitutes the value for
$N_g$ given above and also replaces $N_f$ by $N_f/2$ since quarks and
squarks are in  real representations and thus contribute only half as many
degrees of fredom as in the  $SU(N)$ case.
In the dual magnetic description the gauge group is replaced by $SU(N_f-N_c)$
but the number of flavours remains $N_f$.
Following Seibergs duality hypothesis, we also need
$N_f\times N_f$ free chiral multiplets, which for $SU(N)$adds $2N_f^2$ bosonic
degrees of
freedom (and an equal number fermionic ones), and thus for the dual "magnetic"
description we get
\be
{\it P}_0^{mag}&=&\left[2[(N_f-N_c)^2-1)]+4(N_f-N_c)N_f+2N_f^2\right]
\left(1+\frac7  8\right) \frac{\pi^2T^4}{90}\nonumber\\
&=&P_0^{el} + 4N_f\left(N_f-\frac{3N_c}{2}\right)\frac{\pi^2T^4}{24}
\ \ \ \ \ .
\ee
For SO(N) we have a similar expression remembering that the gauge group is
now $SO(N_f - N_c +4)$ and that the meson multiplet is formed from a
symmetric combination of two complex fields, and thus has
$2\,N_f(N_f+1)/2$ components.

Comparing (2) and (3),  we see that if $N_f<3N_c/2$ the
number of "blackbody" degrees of freedom in the electric description is
greater than the corresponding number in the magnetic description. This is a
reasonable result: The same condition $N_f<3N_c/2$ describes the
the range of $N_f$ for which the magnetic description is infrared free and
the electric theory is confining. (We remind the reader that in the region
$3N_c/2 <N_f< 3N_c$ Seiberg has argued that we should have a non-abelian
Coulomb phase and that the electric theory loses asymptotic freedom at the
upper of these limits.) We
would certainly expect the number of massless "hadrons" to be smaller than the
number of constituents in a confining theory, and this is exactly what we find.
A similar calculation for the case of  $SO(N_c)$, where the limiting case
is when $N_f=3(N_c -2)/2$, gives the same
result:  The number of massles degrees of freedom in the electric theory is
larger than the ones in the dual magnetic theory as long as the electric
theory is confining and the magnetic is infrared free.

\input psfig.tex
\begin{figure}[htb]
\centerline{
\psfig{file=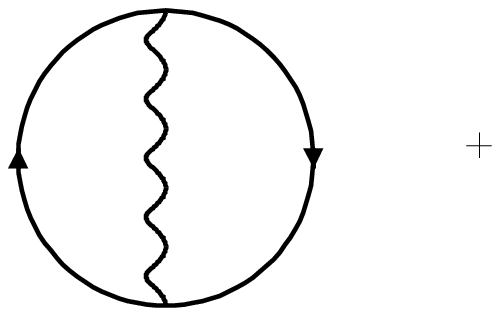,width=1.5in,height=1.5in}
\psfig{file=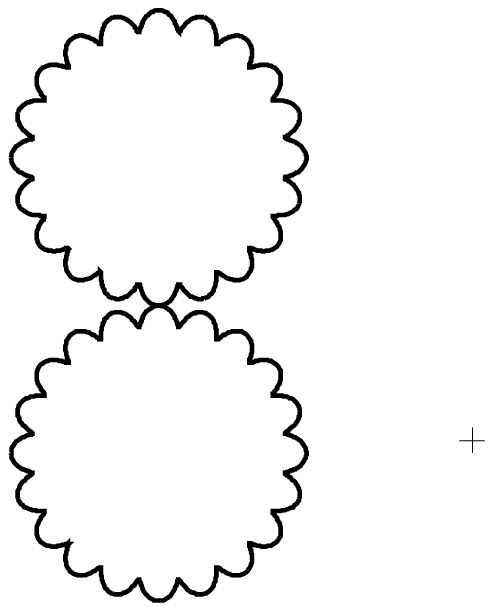,width=1.5in,height=1.5in} }
\caption{Some of the $g^2$-contributions}
\end{figure}

The $O(g^2)$ contribution to the pressure is obtained by
calculating  diagrams of the type shown in fig. 1, using the rules
given in the Appendix. The result is
\be
P_{2}^{el}=-N_g \left(C_A + 3T_f \right)\frac {g^2T^4} {64} \ \ \ \ \ ,
\ee
where $C_A$ is the Casimir operator for the adjoint representation, and
$T_f$ depends on the matter content. For $SU(N)$, $C_A = N_c$ and $T_f =
N_f$, while for $SO(N)$, $C_A = (N_c-2)/2$ and $T_f =N_f/2$.
The
pressure in the magnetic theory is obtained by replacing $N_c$ by
$N_f-N_c$ or $N_f-N_c+4$ for $SU(N)$ and $SO(N)$ respectively.
(Note that the region we consider, the extra chiral fields are
non-interacting, so they contribute only to the blackbody pressure.)

In theories with massless bosons, the next contribution is $O(g^3)$, rather
than
$O(g^4)$, due to the infrared singularities related to the ring diagrams
exemplified in fig. 2. The techniques for calculating these diagrams are
well known, and essentially amonts to calculating the (thermal) electric mass,
$m_{el}$, for the
gluons and the thermal mass, $m_{sq}$, for the squarks. These masses are
readily
obtained from the standard self-energy diagrams, and we find,
\be
m_{el}^2 &=& \half  (C_A+T_f)\, g^2 T^2  \\
m_{sq}^2 &=&  \frac {N_g}{12 N_c}\,g^2 T^2 \nonumber \ \ \ \ \ .
\ee
This implies the following  result for the $O(g^3)$ contribution to the
pressure,
\be
{\it P}_3^{el} =
\left[\left(\frac{C_A+T_f}{2}\right)^{3/2}N_g+2\left(\frac{N_g}
{12N_c}\right)^{3/2}N_cT_f\right]
\frac{g^3T^4}{12\pi} \ \ \ \ \ ,
\ee
$(\frac{N_c}{3}+\frac{N_f}{6})^{3/2}N_g$,
and again the result
in the corresponding magnetic theory is obtained by replacing $N_c$ by
$N_f-N_c$, or by $N_f-N_c+4$.

\begin{figure}[htb]
\centerline{
\psfig{file=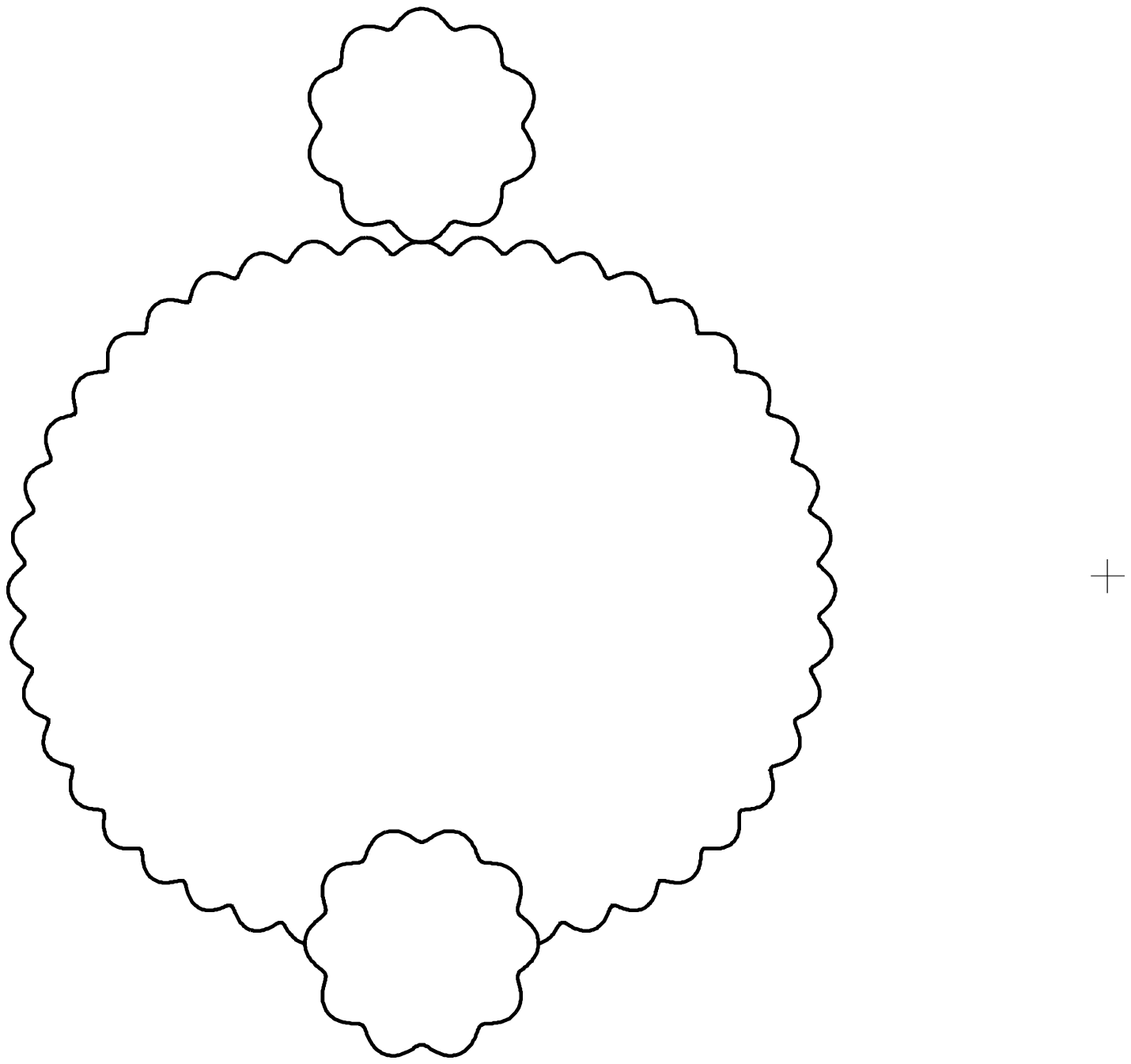,width=1.5 in,height=1.5in}
\psfig{file=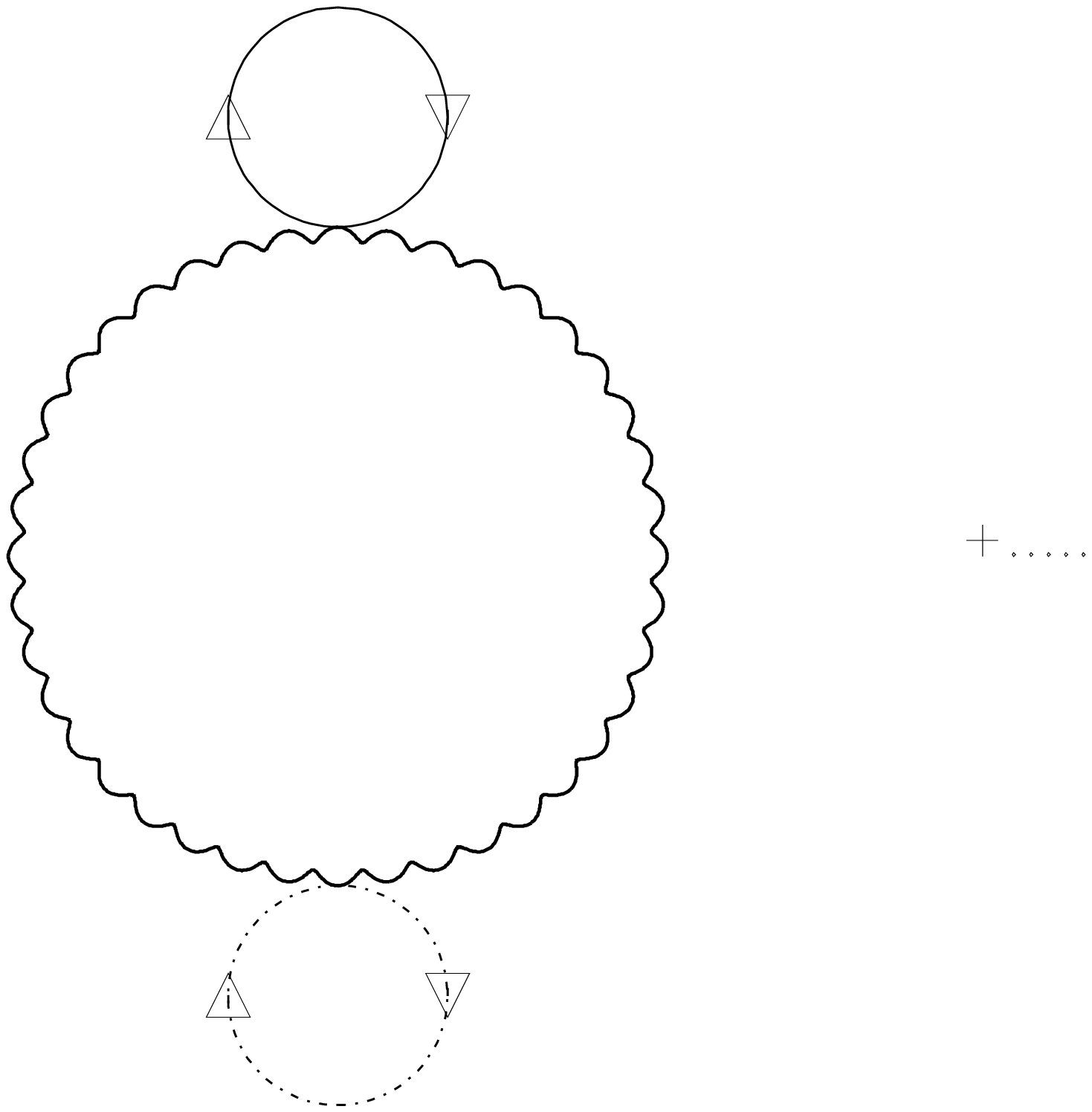,width=1.5in,height=1.5in}   }
\caption{Ring-diagrams}
\end{figure}

To get the total pressure to $O(g^3)$ in the electric theory we add
${\it P}^{el}_0$, ${\it P}^{el}_2$ and ${\it P
}^{el}_3$
and  replace $g^2$ by a runing coupling
constant at temperature T,
\be
g^2_{el} = \frac{8 \pi^2}{b_{el}\ln (T/\Lambda_{el})}
\ee
and for the magnetic theory we add the corresponding results and replace
$g^2$ by
\be
g^2_{mag} =\frac{4\pi^2}{b_{mag} \ln (\Lambda_{mag}/T)}
\ee

For $SU(N)$,  $b_{el}= 3N_c-N_f$ and $b_{mag}= N_f- 3N_c/2 $,
while for $SO(N)$,  $b_{el}= 3N_c/2 - N_f/2 -3$ and $b_{mag} = 3N_c/2-3-N_f$.
The expression for $g^2_{el}$ is valid for $T \gg \Lambda_{el}$ and the one
for $g^2_{mag}$ for $T \ll \Lambda_{mag}$. If we define the ratio
$x=\Lambda_{mag}/\Lambda_{el}$, there is a possibility of using
perturbation theory in both the electric and magnetic theory simultaneously
if $1< T/\Lambda_{el} < x$. This will be discussed further below.
Finally we should stress a very important point: Because of supersymmetry,
the vacuum energy is identically zero, and thus also the pressure at zero
temperature. This is not the case for non-supersymmetric theories like QCD,
where there is a non-zero vacuum pressure (related to the quark an gluon
condensates, and phenomenologically manifested by the so called bag
constant) that is important for understanding the finite $T$ phase
transitions \cite{KAPU}.

\vskip 3mm \noi
\noi{\large\bf 3. Discussion }

We shall first discuss the validity of the high and low T expressions for
the pressure derived in the previous section.
It is known that in QCD, the
perturbative expansion of the ($T=0$ subtracted) free energy  is well
defined only to $O(g^5)$, since at  $O(g^6)$ it becomes sensitive to the
non-perturbatively generated $O(g^2)$ magnetic mass. For a good discussion of
the results to $O(g^3)$ see \cite{KAPU}. Recently the  $O(g^4)$ term in pure
Yang-Mills theory has been calculated by Arnold and Zahi \cite{ARNO}.
The conclusion, so far, is  that
perturbation theory is  reliable only at very high temperatures,
and this is likely to  be true also for the theories we are considering
here. Unfortunately, because of finite size corrections, not even the most
recent   lattice simulations of pure YM theory are good enough to really settle
how well perturbation theory works. For theories with dynamical fermions,
like the one studied here, we may  have to wait long before reliable lattice
results are available. Another point which is relevant for this discussion
is the contribution to the pressure from topologically nontrivial gauge
configurations. In particular, for QCD the contributions from  instantons
have been calculated by Pisarski and Yaffe (for a review see \cite{GROS})
who showed that the instantons become important only at temperatures too low
for perturbation theory to be valid. It would be interesting to repeat this
calculation in the supersymmetric case.

The presence of two scale parameters, $\Lambda_{el}$ and  $\Lambda_{mag}$,
requires some comment. $\Lambda_{el}$ is the usual renormalization group
invariant scale of the asymptotically free (electric) gauge theory, and
just as in ordinary QCD it sets the scale for the onset of nonperturbative
phenomena. While less familiar, the situation is very  similar   in
an IR free theory. The reason that this is not always stressed in the
prototype IR free theory, QED, is that  the electron mass provides a
natural renormalization point.  Since $\alpha$ is very small at this
scale, nonperturbative effects become important only at  distances so short
that  QED must be extended to the full electroweak theory.
If, however, the electron had been massless,
using a running coupling constant and a renormalization group invariant
scale would be as natural in QED as in QCD. Again the renormalization group
invariant scale parameter $\Lambda_{QED}$ would tell us at what scale the
non-perturbative (short distance) physics becomes important.

In the theories
we consider, naturalness implies that there is only one scale connected to
the onset of nonperturbative phenomena. However, this does not mean that
the $\Lambda$ parameter occuring in the UV (or electric) or IR (or
magnetic) description of the theory are necessarily the same. The
$\Lambda$-parameters depends on the renormalization scheme and could well be
different. In our case $g_{el}$ becomes large as $T$ approaches
$\Lambda_{el}$ from above, and $g_{mag}$ becomes large as $T$ approaches
$\Lambda_{mag}$ from below. Thus there are two possibilities:
1) $\Lambda_{mag} \leq \Lambda_{el}$. In this case the temperature
region in between  $\Lambda_{mag}$ and $\Lambda_{el}$ is not even in
principle accessible in perturbation theory. 2) $\Lambda_{mag} >
\Lambda_{el}$. Here there is a $\em possibility$ of simultaneously using
perturbation theory in both the electric and the magnetic description, but
there is of course no guarantee, or even great hope, that the perturbative
expansions will be good in the overlap region. At present we have no way of
determining which of these alternatives is correct - this could only be
settled by matching a direct nonperturbative calcualtion in the electric
theory, \eg using lattice methods, to a corresponding weak coupling
calculation in the magnetic theory. Any concrete idea for how to
perform such a calculation would be very desireable.

\begin{figure}[htb]
\centerline{
\psfig{file=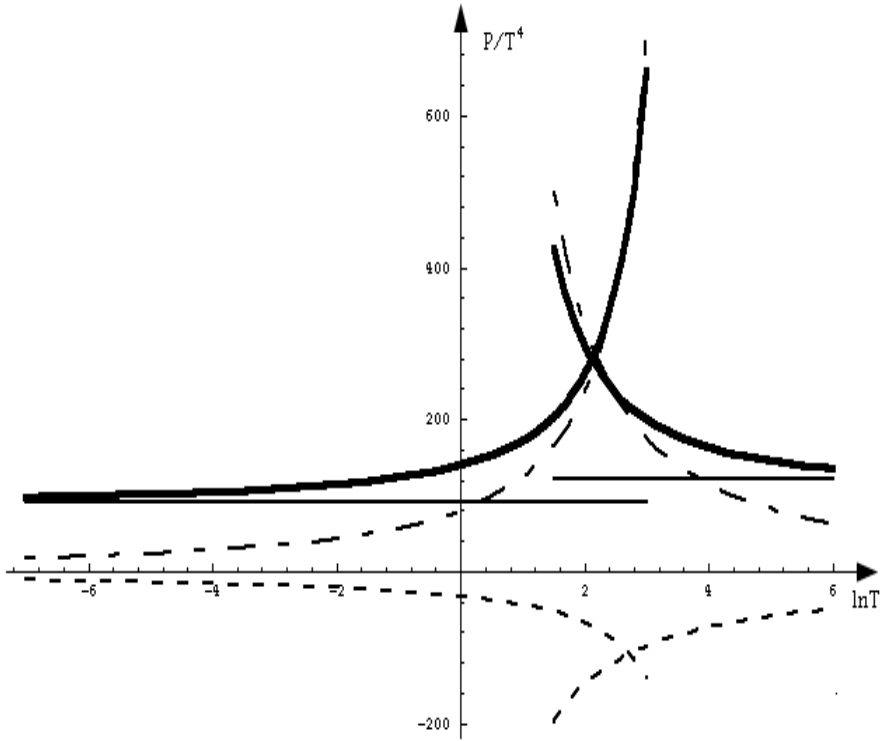,width=12cm,height=10cm}
}
\caption{Pressure versus logaritm of the temperature for $N_c=9$,
$N_f=12$, $\Lambda_{el} = 1$ and
$\Lambda_{mag}=x\Lambda_{el}=50\Lambda_{el}$. The thin solid lines are the
blackbody contributions, the dotted lines the $O(g^2)$, the dashed lines the
$O(g^3)$ and the thick lines the total contributions.  }
\end{figure}
Clearly, the second of the above alternatives is the more interesting one,
and we shall now investigate it a bit further. In fig. 3 we show the $SU(N)$
$O(g^0)$, $O(g^2)$ and $O(g^3)$ contributions to the pressure as a function
of $\ln(T/\Lambda_{el})$ for $x=\Lambda_{mag}/\Lambda_{el}=50$. We see that
even
though
the ratio $x$ is taken (presumably unrealistically) large perturbation theory
looks very poor in the overlap region. The reason is  the large
coefficients in front of the logarithms in the $O(g^2)$ and $O(g^3)$
terms. One might have hoped to see a qualitative difference between the
the $SU(N)$ and  $SO(N)$, since  a phase transition in expected for the
latter, but in fact fig. 3 is typical for both gauge groups.
\begin{figure}[htb]
\centerline{
\psfig{file=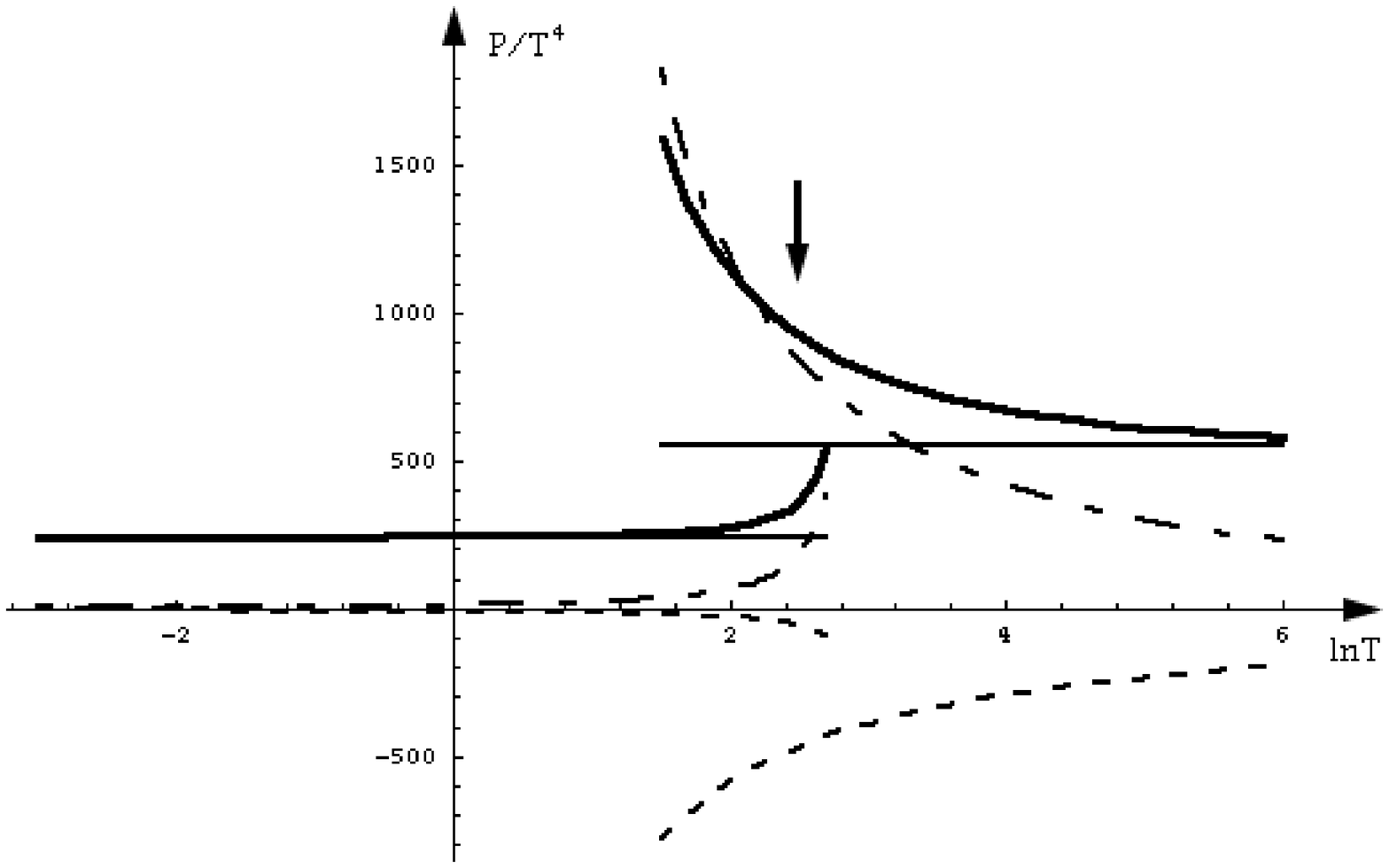,width=12cm,height=10cm}
}
\caption{Pressure versus logaritm of the temperature for $N_c= N_f=30 $,
 $\Lambda_{el} = 1$ and $\Lambda_{mag}=x\Lambda_{el}=20\Lambda_{el}$.
The lines have the same meaning as in fig. 3. }
\end{figure}

An obvious question is whether there is a range of
parameters $N_c$ and $N_f$ where the coefficents of the logarithms become
small.
The answer
is yes, by simply noting that for $N_f$ very close to $N_c$ the number of
magnetic gluons is small and thus also the coupling in the magnetic
theory. This is most clearly seen in the large $N_c$ limit where the
$O(g^2)$ and $O(g^3)$ corrections vanish for $N_f/N_c\rightarrow 1$.
In fig. 4 we show the different components of the pressure in the $SO(N)$
case, where a phase transition in expected, for
$N_c = N_f = 30$ and $x=20$.
In this, admittedly rather extreme case,
it is not unreasonable to assume
that both perturbation expansions can be trusted in the region in the
proximity of the arrow. The shape of the curves are however not
compatible with having a phase transition, where the pressure must be
continuous, and concavity of the free energy density requires that
discontinuities
of the derivative of the pressure, and hence of $P/T^4$, must be such that it
increases with $T$.
{}From this we can conclude that the value $x=20$ cannot be correct, so the
perturbative analysis have in fact allowed us to put a bound on the
intrinsically non-perturbative parameter $\Lambda_{mag}/\Lambda_{el}$. Also
note, that a bound on $x$ also translates into a bound
on $T_c/\Lambda_{el}$, where $T_c$ is the transition
temperature. Clearly these bounds are not very strict, and we
have not made any attempts to systematically explore different
values for $x$, $N_f$ and $N_c$.

In conclusion, we have explored the possibility of using perturbation
theory to study both the  high and low temperature regime in N=1 SUSY
theories possessing Seiberg duality. First we noted that
the number of black-body degrees of freedom in the electric theory is
larger than in the magnetic, exactly in the region where the former is UV
freee and the second IR free, which is what is naively expected in a
confining theory. We calculated the  pressure   to $O(g^3)$ in both
the UV and IR regime, and discussed various possibilities for the
intermediate temperature regime. Our results are somewhat disappointing,
in the sense that the dream of finding a regime where both
perturbation expansions were applicable was not fullfilled. We did, however,
manage to glean some nonperturbative information from our calculations by
studying the large $N_c$ limit.

There are several directions in which this work might be extended that
could be of interest. We have already mentioned that instanton effects
could change our resulsts, and so could of course higher order terms in
perturbation theory. Another interesting question is to study the deformed
theories, either by including mass terms, or by  breaking the global
symmetries by choosing moduli parameters different from zero. In particular
one could ask whether finite temperature effects would induce a flow in
moduli space, and thus making some vacuua unstable against bubble
formation.

\vskip 2mm \noi
{\bf Acknowledgements:  } We thank A. Fayyazuddin, and
P. Di Vecchia for  discussions and  N. Seiberg for a useful correspondence.

\newpage \noi
\noi{\large\bf Appendix }
\renewcommand{\theequation}{A.\arabic{equation}}
\setcounter{equation}{0}
\vskip 2mm\noi
{\em A. Feynman rules}

In components, the lagrangian density (1) takes the form,
\be
{\cal L} =-\frac 1 4 F^{A\mu\nu}F^A_{\mu\nu} +i\lambda ^{\alpha A}
(D\bar{\lambda})_{\alpha A}
+i{\tilde\psi}^{\alpha
a}(D \bar{\tilde{\psi}})_{\alpha a}
-i\bar{\psi}^{\dot{\alpha}a}( D\psi)_{\dot{\alpha}a}
\nonumber\\\nonumber\\
+(D^{\mu}\varphi)^{+a}(D_{\mu}\varphi)_a+
(\tilde{D}^{\mu}\tilde{\varphi})^a
(\tilde{D}_{\mu}\tilde{\varphi})^+_a\nonumber\\ \nonumber\\
+i\sqrt{2}g\left[(\varphi^{\star}T^A\psi^{\alpha})\lambda_{\alpha}^{A}+
(\bar{\psi}^{\dot{\alpha}}T^A\varphi)\bar{\lambda}^A_{\dot{\alpha}}
+\lambda^A_{\alpha}(\tilde{\psi}^{\alpha}T^A\tilde{\varphi}^{\star})
+(\tilde{\varphi}T^A\bar{\tilde{\psi}}_{\dot{\alpha}})
\bar{\lambda}^{\dot{\alpha}A}\right ]\nonumber\\ \nonumber\\
-\frac 1 2 g^2 \sum_{a}\left[(\varphi^{\star}T^A\varphi)-
(\tilde{\varphi}T^A\tilde{\varphi}^{\star})\right]^2
\ee
where the matrices $T^A$, $A=1...N_g$ span the fundamental
representation of $SU(N_c)$, and where we have suppressed contractions of
group indices. The various covariant derivatives are defined as follows:
\be
(D\bar{\lambda})_{\alpha A}&=&(\sigma^{\mu})_{\alpha\dot{\alpha}}
[\partial _{\mu}\delta_{AB} + gf_{ABC}A^C_{\mu}]\lambda^b
\nonumber\\\nonumber\\
(D\psi)^{\dot{\alpha}}_{a}&=&(\bar{\sigma}^{\mu})^{\dot{\alpha}\alpha}
[\partial _{\mu}\delta_a^b +igA^A_{\mu}(T^A)_a^b]\psi_{\alpha b}
\nonumber\\\nonumber\\
(D\bar{\tilde{\psi}})_{\alpha a}&=&(\sigma^{\mu})_{\alpha\dot{\alpha}}
[\partial_{\mu}\delta_a^b
+igA^{A}_{\mu}(T^A)_a^b]\bar{\tilde{\psi}}^{\dot{\alpha}}_b
\nonumber\\\nonumber\\
(D_{\mu}\varphi)_a&=&[\partial_{\mu}\delta_a^b+igA^A_{\mu}(T^A)^b_a]\varphi_b
\nonumber\\\nonumber\\
(D_{\mu}\varphi)^{+a}&=&\partial_\mu\varphi^{\star a}-ig\varphi^{\star
b}(T^A)_b^rA_{\mu}^A
\nonumber\\\nonumber\\
(\tilde{D}_{\mu}\tilde{\varphi})^a&=&\partial_{\mu}\tilde{\varphi}^a
-ig\tilde{\varphi}^b(T^A)_b^rA^A_{\mu}\nonumber\\\nonumber\\
(\tilde{D}_{\mu}\tilde{\varphi})^+_a&=&[\partial_{\mu}\delta^b_a
+igA^A_{\mu}(T^A)_a^b]\tilde{\varphi}^{\star}_b
\ee
Our  Minkowski metric is $\eta_{\mu\nu}=diag(1,-1,-1,-1)$,
and the $2\times2$-matrices $\sigma^{\mu}$
and $\bar{\sigma}^{\mu}$ satisfy the relations
\be
\sigma^{\mu}_{\alpha\dot{\alpha}}\bar{\sigma}^{\nu\dot{\alpha}\beta}+
\sigma^{\nu}_{\alpha\dot{\alpha}}\bar{\sigma}^{\mu\dot{\alpha}\beta}=
2\eta^{\mu\nu}\delta_{\alpha}^{\beta}\nonumber\\\nonumber\\
\bar{\sigma}^{\mu\dot{\alpha}\alpha}\sigma^{\nu}_{\alpha\dot{\beta}}+
\bar{\sigma}^{\nu\dot{\alpha}\alpha}\sigma^{\mu}_{\alpha\dot{\beta}}=
2\eta^{\mu\nu}\delta^{\dot{\alpha}}_{\dot{\beta}}
\ee
{}From the action we read off the
Feynman rules given in figs. 6 - 9. For the gauge group $SO(N)$ the rules
are identical except that since the matter multiplets are real, there are
no dotted quark or squark lines. In the calculations, this only amounts to
an overall factor of 2 for the relevant diagrams.

\vskip 2mm\noi
{\em B. Sample Calculation}

We illustrate the calculation of the $g^2$ contributions by evaluating the
diagram in fig. 6 involving one gluon and two gluinos.
\begin{figure}[htb]

\centerline{
\psfig{file=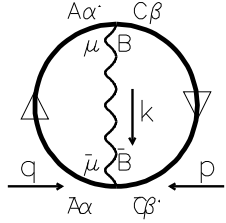,width=1.5in,height=1.5in} }
\caption{Diagram contributing to the pressure to $O(g^2)$. }
\end{figure}

Using the Feynman rules and dropping a temperature independent
piece\cite{KAPU} we find,
\begin{eqnarray}
{\cal I}_{Ggg}&=&-{\textstyle{1 \over
2}}g^2\int_-dpdq\int_+dk\left\{{\left({{i(\bar\sigma
p)^{\dot\beta\beta}\delta^{C\bar C}}\over
p^2}\right)
\left(-(\sigma^\mu
)_{\beta\dot\alpha}f^{ABC}\right)
\left(-{{i(\bar\sigma q)^{\dot\alpha\alpha}\delta^{A\bar A}}\over
q^2}\right)}\right.\cr
&&\cr
&&
\left.{
\left(-(\sigma^{\bar\mu})_{\alpha\dot\beta}f^{\bar C\bar B\bar
A}\right)
\left(-{{i\eta_{\mu\bar\mu}\delta^{B\bar
B}}\over k^2}\right)}\right\}(2\pi)^4\delta(p+q+k)\cr
&&\cr
&=&2ig^2N_cN_g\int_-dpdq\int_+dk\left\{{{pq}\over{p^2q^2k^2}}\right\}
(2\pi)^4\delta(p+q+k),
\end{eqnarray}
where +(-) denotes the momentum integral for a boson (fermion),
cf. (\ref{Trix}) below.
The second equality follows directly using $\sigma$-matrix algebra,
and taking traces. To see that this
integral factorizes we use the delta function to write
$2pq=k^2-p^2-q^2$ and obtain
\begin{eqnarray}
{\cal I}_{Ggg}=ig^2N_cN_g\left(A_-^2-2A_+A_-\right)=i{\textstyle{5\over
4}}g^2N_cN_gA_+^2 \ \ \ \ \ ,
\end{eqnarray}
where,
\begin{eqnarray}
A_\pm\equiv\int_\pm {dp\over p^2}\equiv \mp {1\over{2\pi
i}}\int{{d^3p}\over {(2\pi)^3}}\int_{-i\infty +\varepsilon}^{i\infty
+\varepsilon}dp_0{1\over{e^{\beta p_0}\mp 1}}
\left({2\over{p_0^2-{\underline{p}}^2}}\right)\ \ \ \ \ .\label{Trix}
\end{eqnarray}
Evaluating these integrals yields,
\begin{eqnarray}
A_+=-2A_-=-{T^2\over 12}.\label{Trex}
\end{eqnarray}

\bigskip
Notice the factor $i$ in (\ref{Trex}). It enters because we use Minkowski
space Feynman rules to extract the free energy. A proper translation back to
Euclidean space shows that the result of a $L$-loop diagram evaluated with
the propagators and vertices in figs. 6 - 9 and integrations as in the
above example, has to be multiplied with $(-i)^{L-1}$ to give the proper
contribution to the free energy.

The above example is typical; all our $g^2$-integrals factorize
in this manner and are expressible in terms of $A_\pm$. For the
ring diagrams that yield the $g^3$ contributions, (fig. 2), we must
calculate the self-energy corrections to the bosonic propagators with
zero momentum on the external legs. This last condition again makes
the calculation easy and the result can be expressed in terms of
$A_\pm$. Again, since we work in Minkowski space, some care is needed when
extracting the thermal masses from the Greens functions, in order to get
the correct phase.

\begin{figure}[htb]
\centerline{
\psfig{file=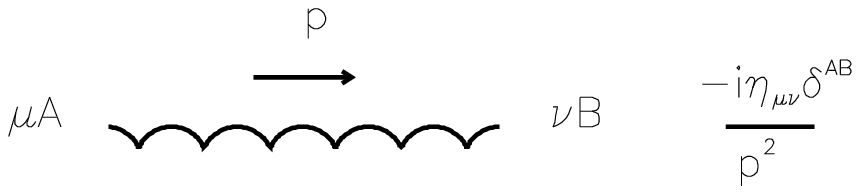,width=3in,height=1.5in}
\psfig{file=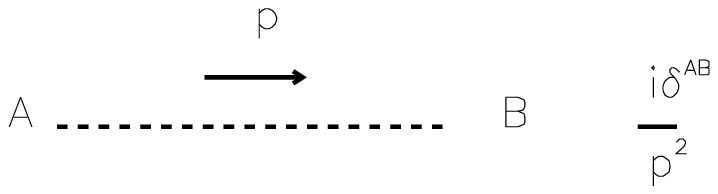,width=3in,height=1.5in}}
\centerline{
\psfig{file=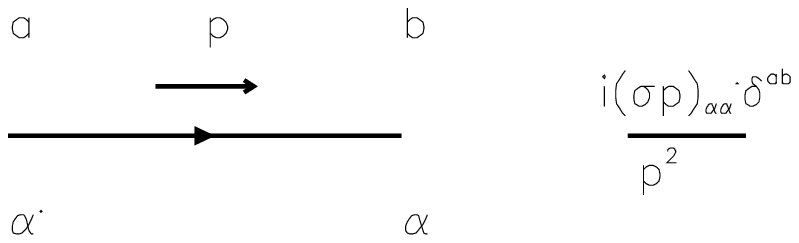,width=3in,height=1.5in}
\psfig{file=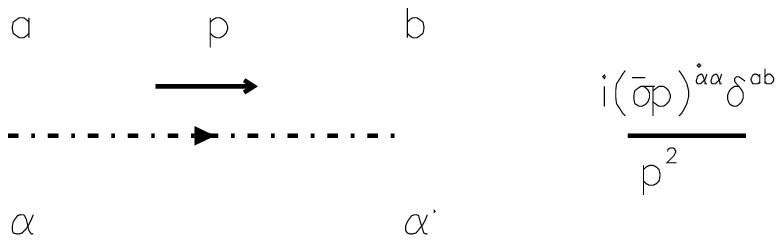,width=3in,height=1.5in}}
\centerline{
\psfig{file=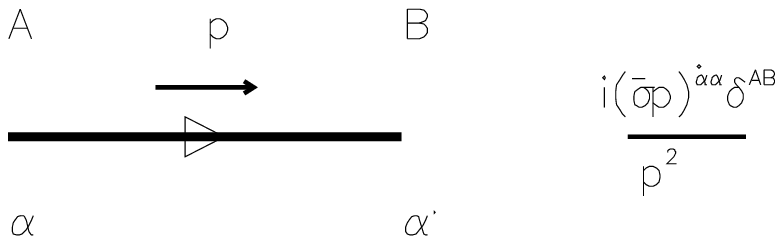,width=3in,height=1.5in}}
\centerline{
\psfig{file=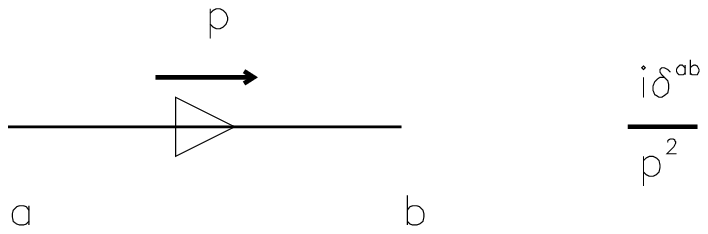,width=3in,height=1.5in}
\psfig{file=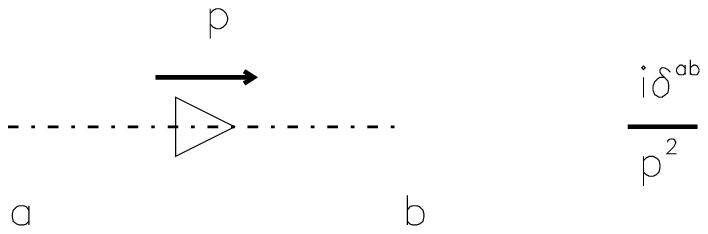,width=3in,height=1.5in}}

\caption{Propagators for {\it{ gluon, ghost, right quark, left quark, gluino,
right squark}} and {\it{left squark}}, respectively. }
\end{figure}

\begin{figure}[htb]
\centerline{
\psfig{file=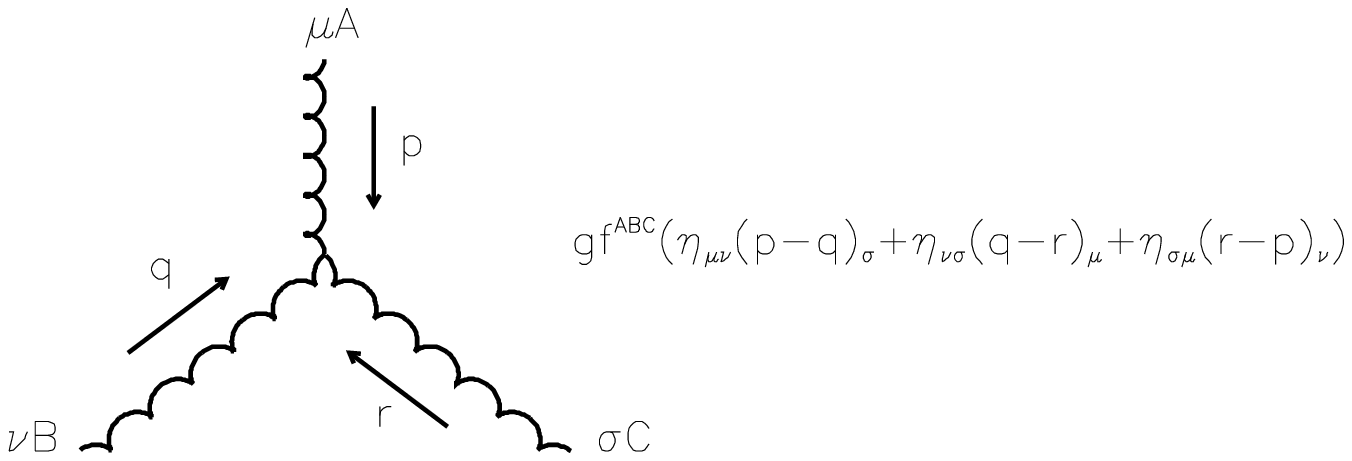,width=2in,height=2in}}
\centerline{
\psfig{file=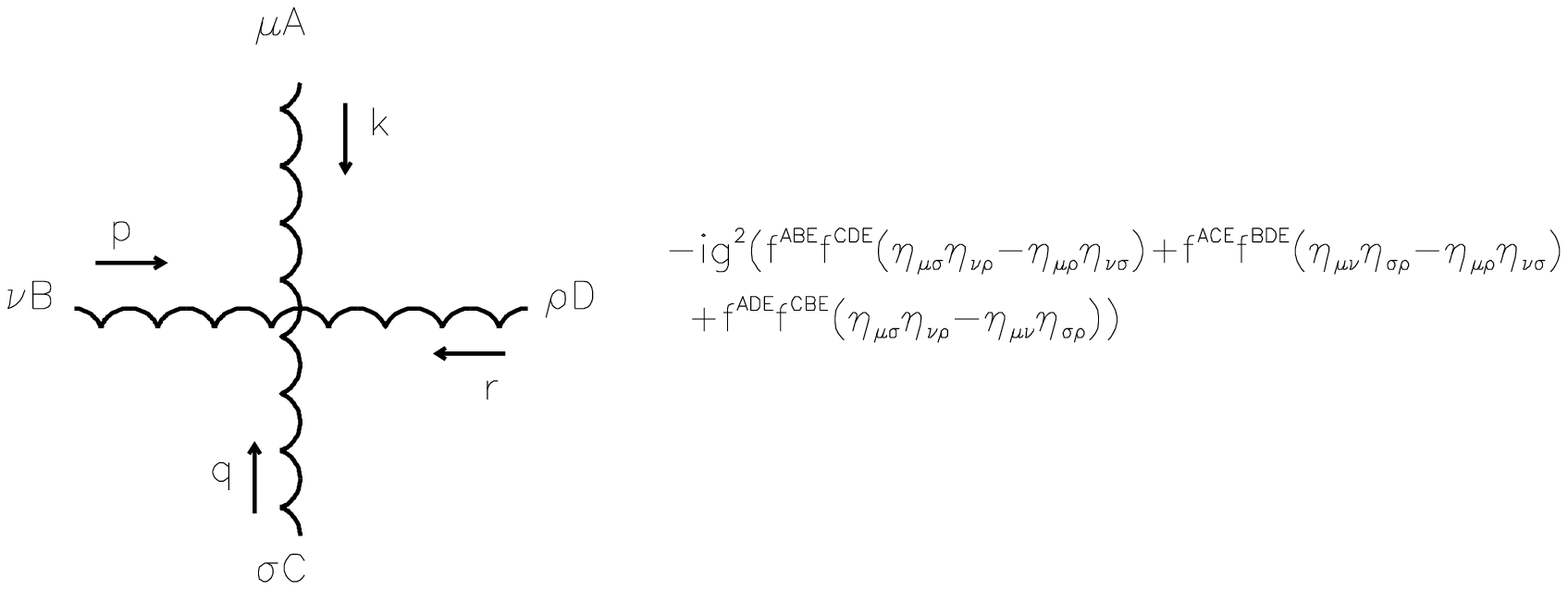,width=2in,height=2in}}
\centerline{
\psfig{file=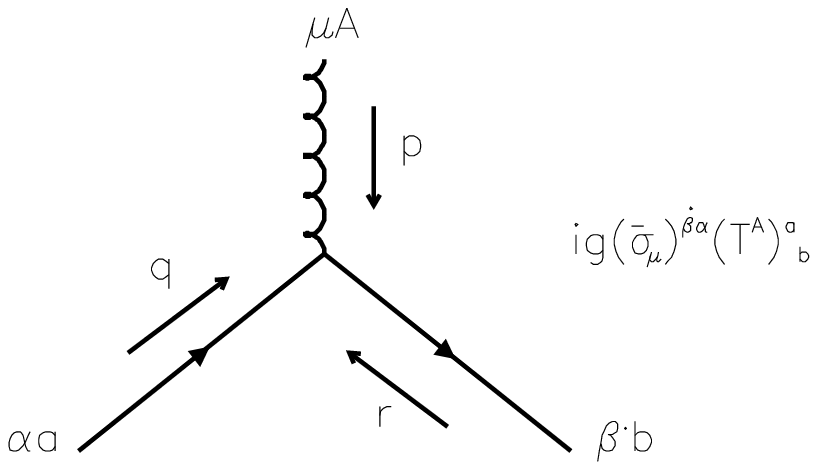,width=2in,height=2in}
\psfig{file=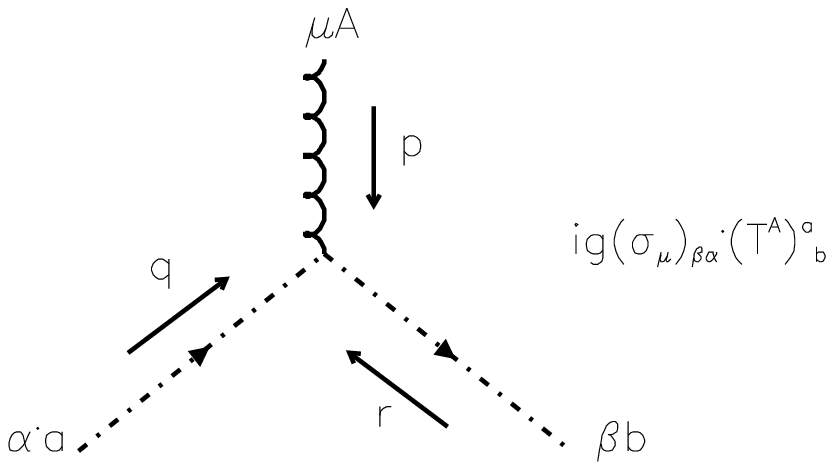,width=2in,height=2in}}
\centerline{
\psfig{file=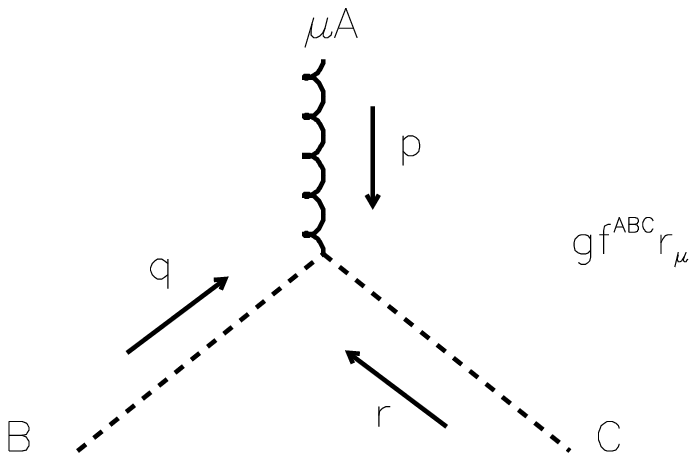,width=2in,height=2in}
\psfig{file=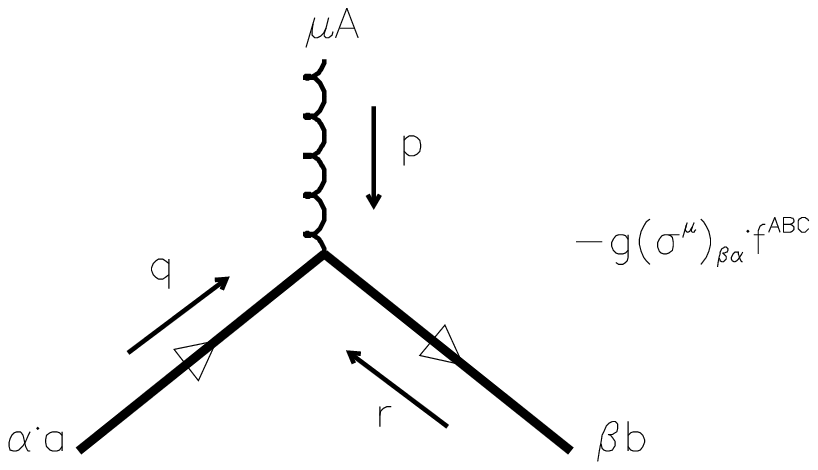,width=2in,height=2in}}

\caption{{\it{Gluon-Fermion}} Vertices}
\end{figure}

\begin{figure}[htb]
\centerline{
\psfig{file=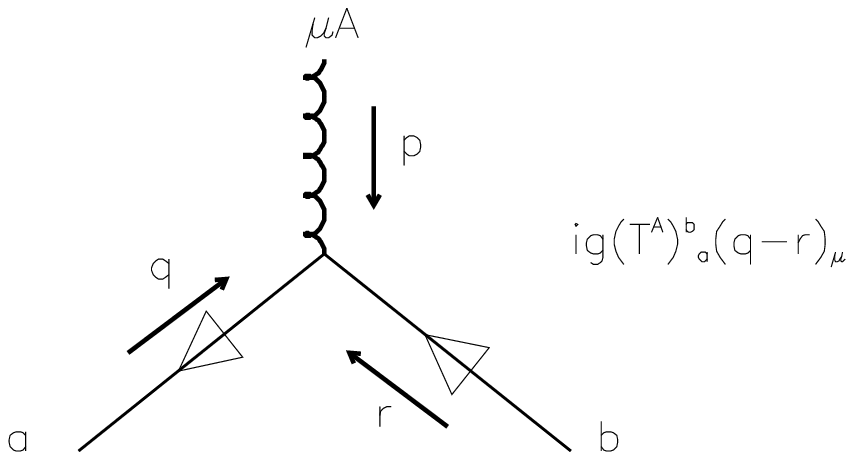,width=2in,height=2in}
\psfig{file=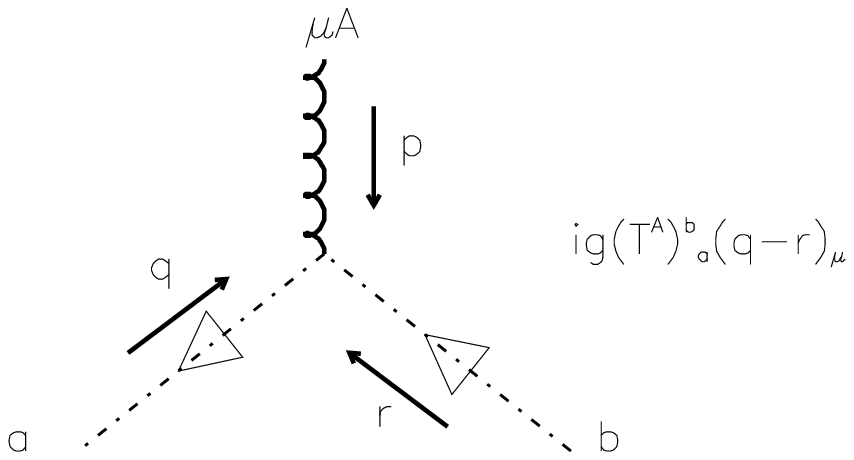,width=2in,height=2in}}
\centerline{
\psfig{file=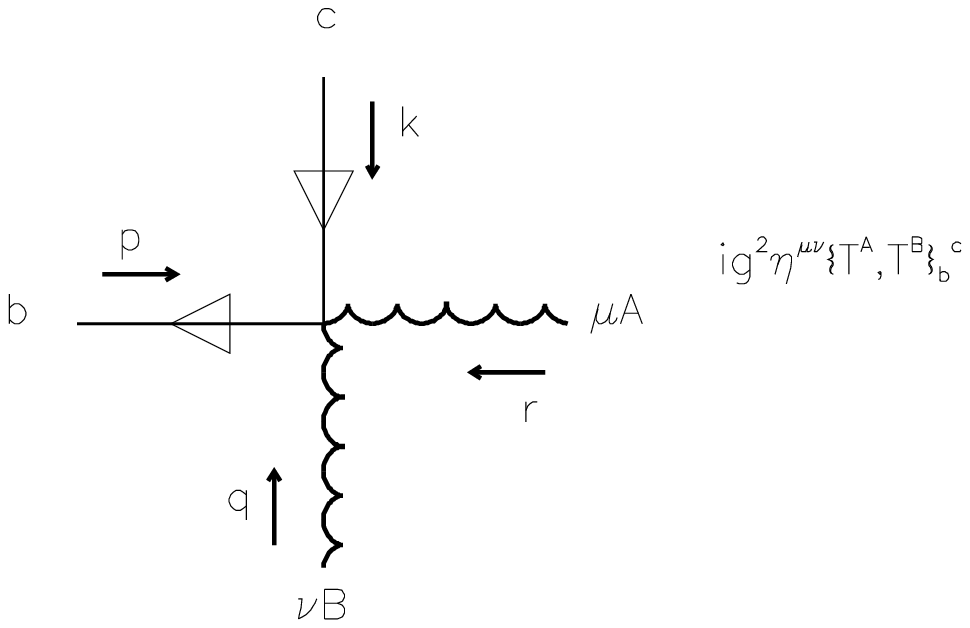,width=2in,height=2in}}
\centerline{
\psfig{file=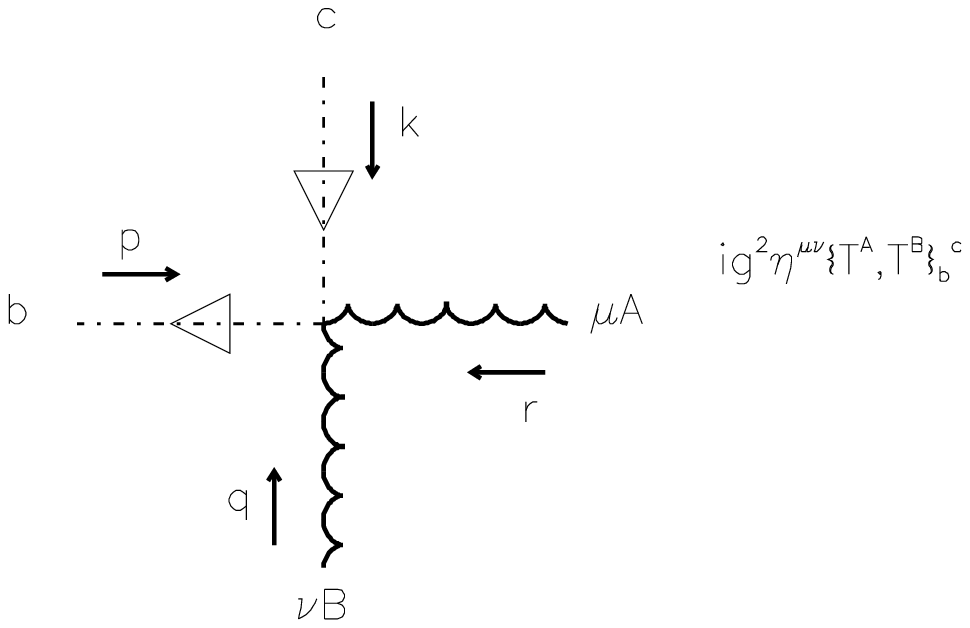,width=2in,height=2in}}
\centerline{
\psfig{file=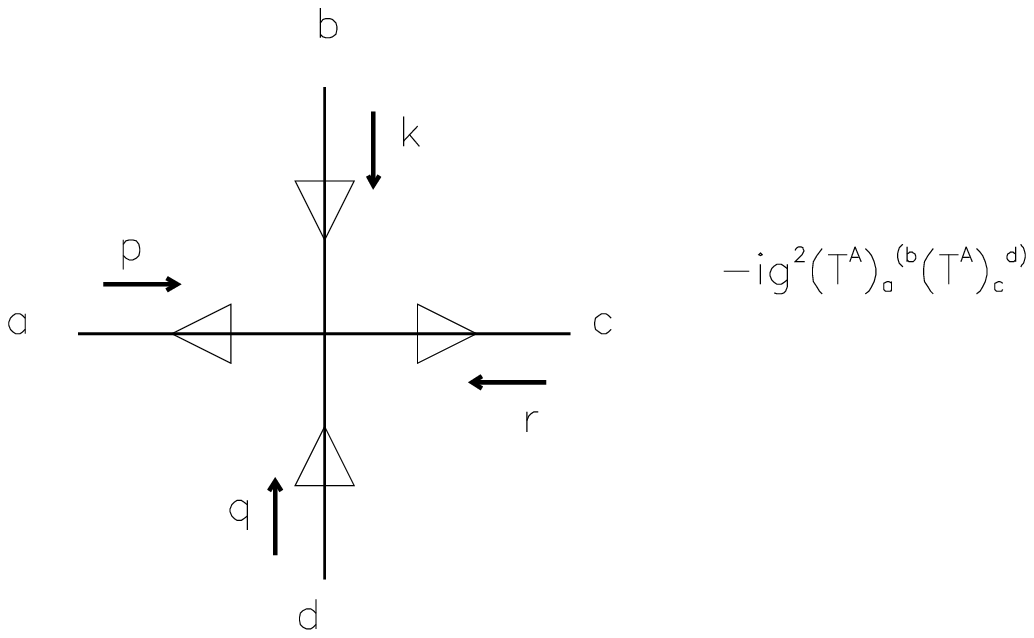,width=2in,height=2in}
\psfig{file=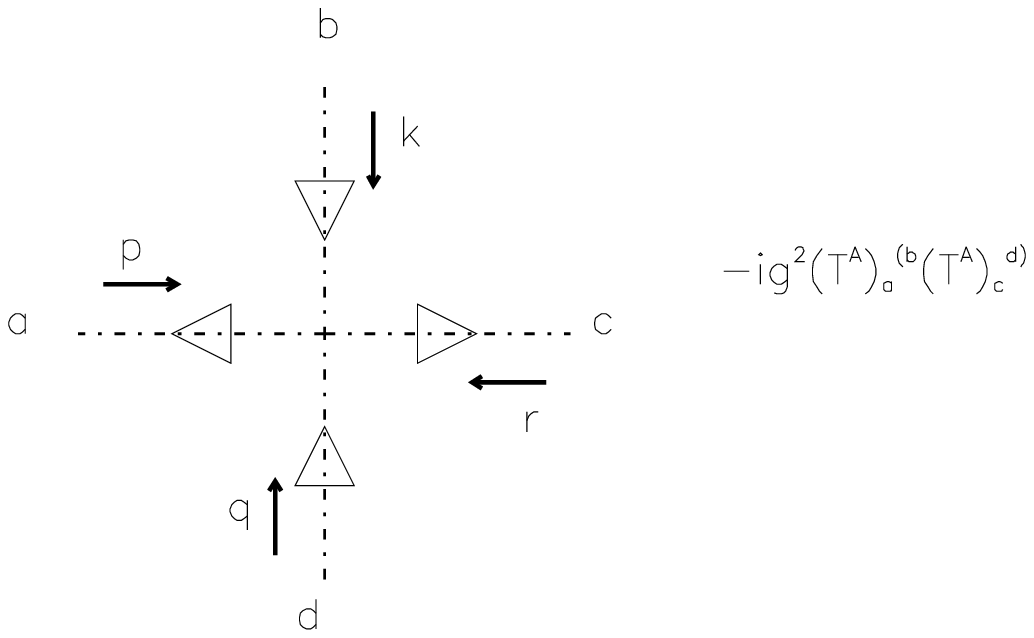,width=2in,height=2in}}
\caption{{\it{Gluon-Squark}} and {\it{Squark-Squark}} Vertices}
\end{figure}

\begin{figure}[htb]
\centerline{
\psfig{file=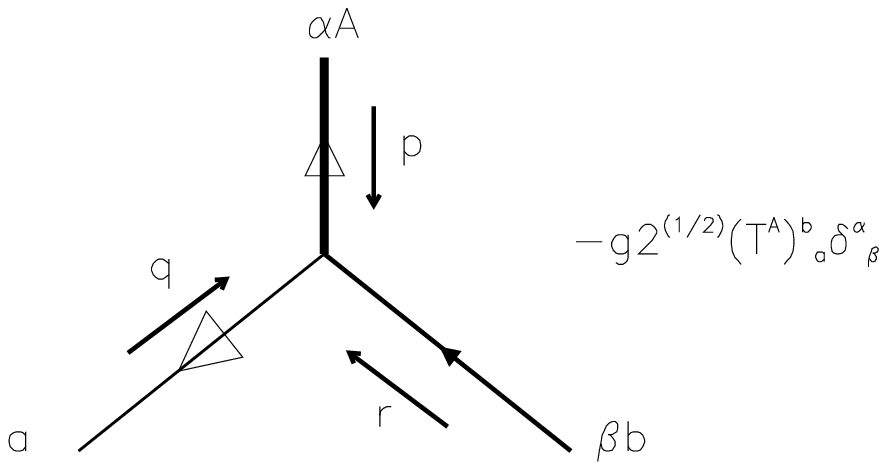,width=2in,height=2in}
\psfig{file=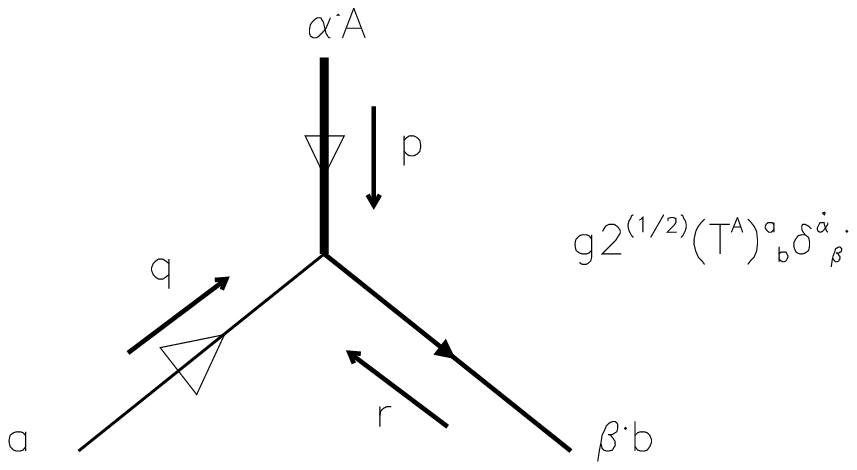,width=2in,height=2in}}
\centerline{
\psfig{file=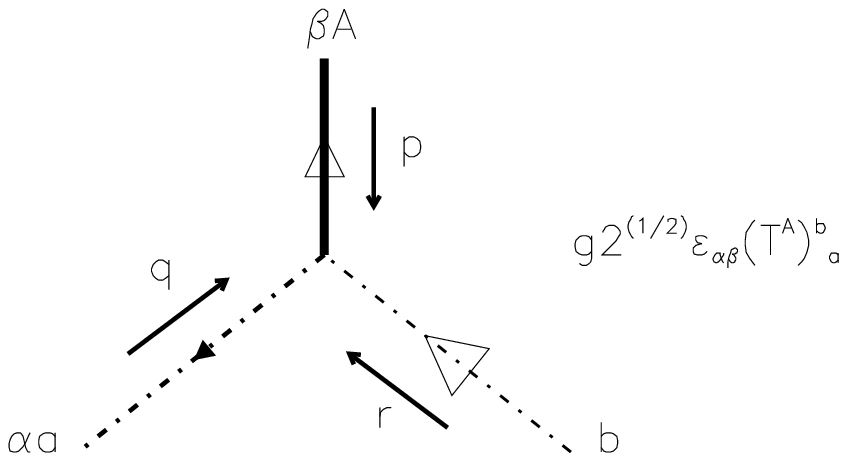,width=2in,height=2in}
\psfig{file=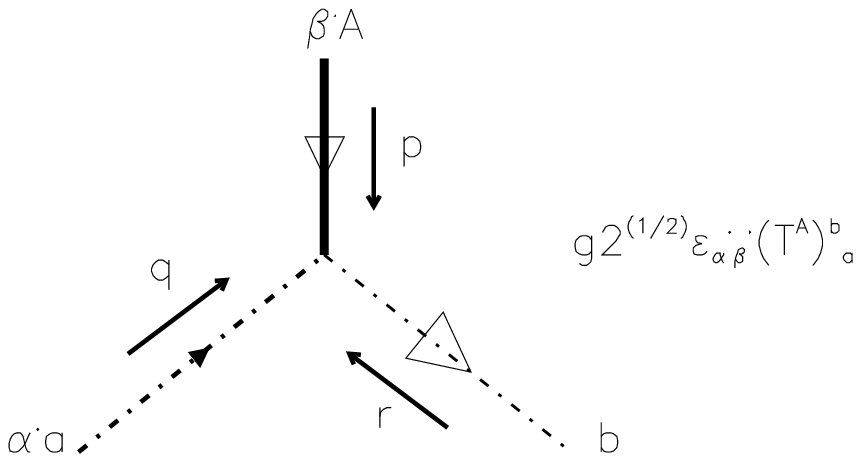,width=2in,height=2in}}
\caption{{\it{Squark-Fermion}} Vertices}
\end{figure}

\end{document}